\documentclass[fleqn,10pt]{wlscirep}
\usepackage[utf8]{inputenc}
\usepackage[T1]{fontenc}
\usepackage{bm}
\usepackage{comment}

\title{Reconfigurable magnonic mode-hybridisation and spectral control in a bicomponent artificial spin ice}

\author[1,3,*]{Jack C. Gartside}
\author[1,3]{Alex Vanstone}
\author[1,2]{Troy Dion}
\author[1]{Kilian D. Stenning}
\author[2]{Daan M. Arroo}
\author[2]{Hide Kurebayashi}
\author[1]{Will R. Branford}
\affil[1]{Blackett Laboratory, Imperial College London, London SW7 2AZ, United Kingdom}
\affil[2]{London Centre for Nanotechnology, University College London, London WC1H 0AH, United Kingdom}
\affil[3]{These authors contributed equally}
\affil[*]{Corresponding author e-mail: j.carter-gartside13@imperial.ac.uk}
\begin{abstract}

Strongly-interacting nanomagnetic arrays are finding increasing use as model host systems for reconfigurable magnonics. The strong inter-element coupling allows for stark spectral differences across a broad microstate space due to shifts in the dipolar field landscape. While these systems have yielded impressive initial results, developing rapid, scaleable means to access a broad range of spectrally-distinct microstates is an open research problem.

We present a scheme whereby square artificial spin ice is modified by widening a `staircase' subset of bars relative to the rest of the array, allowing preparation of any ordered vertex state via simple global-field protocols. Available microstates range from the system ground-state to high-energy `monopole' states, with rich and distinct microstate-specific magnon spectra observed. Microstate-dependent mode-hybridisation and anticrossings are observed at both remanence and in-field with dynamic coupling strength tunable via microstate-selection. Experimental coupling strengths are found up to $g / 2 \pi = 0.15~ \text{GHz}$. Microstate control allows fine mode-frequency shifting, gap creation and closing, and active mode number selection.
\end{abstract}

\begin{document}

\flushbottom
\maketitle
\thispagestyle{empty}

\section*{Introduction}

The field of magnonics aims to employ spin-waves to propagate and process information\cite{kruglyak2010magnonics,lenk2011building}. Spin-waves offer a host of attractive benefits as data carriers including low heat generation, power consumption\cite{chumak2015magnon} and coherent coupling to photons\cite{tabuchi2014hybridizing}, phonons\cite{zhang2016cavity} and other magnons\cite{kalinikos1986theory,liensberger2019exchange,shiota2020tunable,sud2020tunable}. Functional magnonics has proliferated in recent years, with wide-ranging applications from transistors\cite{chumak2014magnon} to multiplexers\cite{vogt2014realization} and logic gates\cite{stenning2020magnonic}. As the complexity of magnonic designs increases, so does the demand for versatile, reconfigurable host systems. 

\begin{figure}[thbp]
\centering
\includegraphics[width=0.7\textwidth]{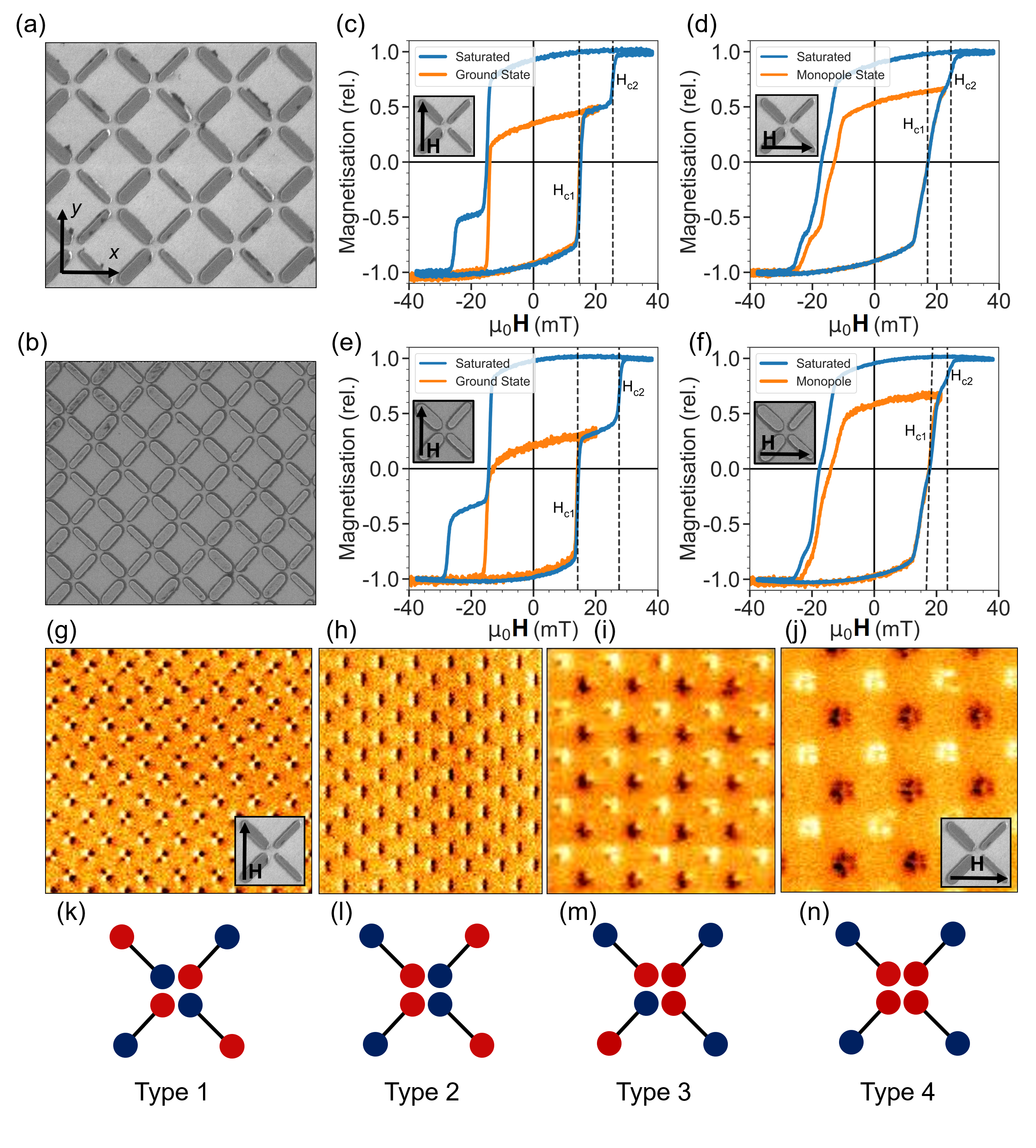}
\caption{Schematic of width-modified square and high-density square reconfigurable magnonic crystals and their type 1-4 microstates and hysteresis loops. $y$ and $x$ array axes are referred to as `ground-state' and `monopole' orientations throughout this work.\\
a) Scanning electron micrograph of the square sample. Bars are 830 nm long, 230 nm (wide-bar) and 145 nm (thin-bar) wide, 20 nm thick with 120 nm vertex gap (bar-end to vertex-centre). \\
b) Scanning electron micrograph of the high-density square sample. Bars are 600 nm long, 200 nm (wide-bar) and 125 nm (thin-bar) wide, 20 nm thick with 100 nm vertex gap. \\
c-f) MOKE hysteresis loops of S sample in `ground state' (c) and `monopole' (d) orientations, HDS sample in `ground state' (e) and `monopole' (f) orientations. Blue points show fully-saturating hysteresis loop, orange points show minor loops with maximum positive field value chosen to prepare sample in type 1 (c,e), type 4 (d) and type 3 (f) states before sweeping back to negative saturation.\\
g-j) Magnetic force microscope images of type 1-4 microstates. Type 1 and 4 states have inset SEM images showing the relative orientation of $\mathbf{H}_{ext}$ to the width-modified subsets required for state preparation. Type 2 and 3 states may be prepared in either $\pm90^{\circ}$ field orientation. Type 1 and 4 states are often termed `ground state' and `monopole' state in artificial spin ice. \\
k-n) Magnetic charge dumbbell schematic of type 1-4 microstates.
}
\label{Fig1} 
\end{figure}

Recently, a family of metamaterials termed reconfigurable magnonic crystals (RMC)\cite{grundler2015reconfigurable,topp2010making,krawczyk2014review,haldar2016reconfigurable,barman2020magnetization} has made strong progress in answering this need. Typically comprising discrete nanopatterned magnetic elements closely-packed in arrays to promote strong dipolar coupling, RMC support multiple microstates and exhibit distinct microstate-dependent magnonic dynamics and spectra with diverse functional benefits. A subset of RMC has emerged based on artificial spin ice (ASI) arrays\cite{wang2006artificial,lendinez2019magnetization,gliga2020dynamics,skjaervo2020advances,barman2020magnetization,talapatra2020magnetic} where geometrical frustration gives rise to a vastly degenerate microstate space that features a long-range ordered ground state \cite{morgan2011thermal,gartside2018realization} and high-energy `magnetic monopole'-like excited states\cite{ladak2010direct,mengotti2011real}. The potential to leverage these states for their magnonic properties is great and studies into fundamentals of state-spectra correspondence\cite{gliga2013spectral,zhou2016large,arroo2019sculpting,dion2019tunable,iacocca2016reconfigurable,bang2020influence,sklenar2013broadband} have set the scene for a new generation of ASI-based RMC designs. An open problem in the field is developing reliable, versatile and rapid means for microstate access. While ASI possesses a huge range of states, they remain largely unavailable for magnonic exploitation due to state preparation techniques being overly simplistic (for example global-field protocols which may only prepare saturated or randomly demagnetised, unrepeatable states), overly slow (surface-probe microscope nanomagnetic writing techniques\cite{gartside2016novel,gartside2018realization,wang2016rewritable,lehmann2019poling} which may prepare any state but on timescales unsuitable for technological integration) or difficult to realise with current nanofabrication techniques (for example, recently proposed multi-level stripline technique\cite{gartside2020current}). In the absence of such techniques, ASI systems have been modified to allow access to an enhanced microstate range using global fields, `magnetic charge ice' which rotates a subset of bars in square ASI to allow global-field preparation of three microstates\cite{wang2016rewritable,wang2018switchable,iacocca2020tailoring} (types 1-3 as seen in fig. \ref{Fig1}) or bar subset modification via either material\cite{lendinez2020emergent}, shape-anisotropy\cite{dion2019tunable} or exchange-bias\cite{parakkat126configurable}. The magnetic charge ice case is elegant, but the way in which bars are rotated leads to greater separations between neighbouring elements so that greater density is required to achieve an appreciable interelement coupling required for collective excitations. Moreover, the rotation means different sublattices will in general experience different effective global excitation and bias fields.

Here, we present a square ASI with a `staircase'-pattern subset of width-modified bars. Shown in figure \ref{Fig1}, this enables preparation of four distinct type 1-4 microstates (fig. \ref{Fig1} b-e), g-j) including the typically elusive ground-state (type 1) and `monopole' states (types 3 and 4). The four states display rich and distinct magnonic spectra with fine control over mode frequency shifting, gap opening and tuning, and number of active modes. Microstate-dependent mode-hybridisation and anticrossings are observed with coupling-strength and gap width tunable via state selection. Selective mode-hybridisation offers reconfigurable mode profile and index control in-field and crucially at remanence.

\section*{Results and Discussion}

\subsection*{Microstate access via width-modification}

Samples were designed by taking square ASI and increasing the width of one nanobar subset. Two samples were fabricated, square (S, fig. \ref{Fig1} a) and high-density square (HDS, fig. \ref{Fig1} b), using an electron beam lithography liftoff process and thermal deposition of Ni$_{81}$Fe$_{19}$. Sample dimensions were selected such that the wider bar subset may be magnetically reversed via a global-field without thin-bars also reversing from the combination of global-field $\mathbf{H}_{ext}$ and local dipolar field $\mathbf{H}_{loc}$, satisfying $\mathbf{H}_{c2} > \mathbf{H}_{ext} + \mathbf{H}_{loc} > \mathbf{H}_{c1}$ with $\mathbf{H}_{c1}$ and $\mathbf{H}_{c2}$ the wide and thin-bar coercive fields respectively ($\mathbf{H}_{c1}$ and $\mathbf{H}_{c2}$ visible in fig. \ref{Fig1} MOKE loops). This enables preparation of the entire range of ordered, `pure' (i.e. single vertex type across the array) microstates. Arrays comprising identical bars may only access a single pure microstate by saturating with global-field (type 2). The width-modification employed here allows global-field access to four pure states with distinct local-field profiles and corresponding magnonic spectral dynamics. Microstates are shown via magnetic force microscope (MFM) (fig. \ref{Fig1} g-j) and magnetic charge schematics (fig. \ref{Fig1} k-n). The S sample comprises bars of 830 nm $\times$ 230 nm (wide-bar), 145 nm (thin-bar) $\times$ 20 nm, 120 nm vertex gap (defined bar end to vertex centre). The HDS sample comprises bars of 600 nm $\times$ 200 nm (wide-bar), 125 nm (thin-bar) $\times$ 20 nm, 100 nm gap.

Bars are widened in alternating $y$-axis columns (axes defined in fig. \ref{Fig1} a) such that wide-bars may be reversed from a saturated background state (type 2, fig. \ref{Fig1} h,l) without reversing thin-bars. If the global-field $\mathbf{H}_{ext}$ is oriented along the $y$-axis, reversing only wide-bars from a $\hat{y}$-saturated state leaves the system in the antiferromagnetic type 1 state (fig. \ref{Fig1} g,k), which forms the ASI ground state with and without width-modification\cite{morgan2011thermal,nisoli2007ground,moller2009magnetic}. Here the microstate allows the maximum amount of inter-bar dipolar flux-closure and lowest system energy. If $\mathbf{H}_{ext}$ is oriented along the $x$-axis, reversing wide-bars results in the type 4 state (fig. \ref{Fig1} j,n), termed a `monopole' or `all-in, all-out' state\cite{castelnovo2008magnetic,moller2009magnetic,farhan2019emergent,perrin2016extensive} with four like-polarity magnetic charges at each vertex, highly-repulsive inter-bar dipolar field interactions and maximum system energy. If $\mathbf{H}_{ext}$ has any angular misalignment from the width-modified columns, one of the $\pm45^{\circ}$ wide-bars will experience a higher field along its easy-axis, resulting in that bar reversing at lower $\mathbf{H}_{ext}$. The resulting state with just one of the $\pm45^{\circ}$ wide-bars reversed is the type 3 state (fig. \ref{Fig1} i,m) with three like-polarity and one opposite polarity magnetic charge per vertex. In experiment there is always some angular misalignment and the array will transition between states 2 and 4 via state 3. The field window in which state 3 exists may be broadened by deliberately increasing angular misalignment.
The S array may access type 1-4 states, the HDS array may access states 1-3 but not 4 as the increased $\mathbf{H}_{loc}$ magnitudes arising from smaller inter-bar separation leads to spontaneous reversal of thin-bars from a thin-bar majority charge type 3 to a wide-bar majority type 3 when attempting state 4 access, i.e. $\mathbf{H}_{ext} + \mathbf{H}_{loc} > \mathbf{H}_{c-thin}$.

We analyse mode frequencies following the Kittel equation\cite{kittel1948theory} $f = \frac{\mu_0 \gamma}{2 \pi} \sqrt{\mathbf{H}(\mathbf{H}+\mathbf{M})}$ in the k=0 limit applicable to this work, with $\gamma$ the gyromagnetic ratio and $\mathbf{H} = \mathbf{H}_{ext} + \mathbf{H}_{loc}$. The local dipolar field landscape varies greatly between microstates, with resulting distinct microstate-dependent magnon spectra.

\subsection*{Microstate-dependent magnonic spectra}

\begin{figure*}[thbp]   
\centering
\includegraphics[width=0.99\textwidth]{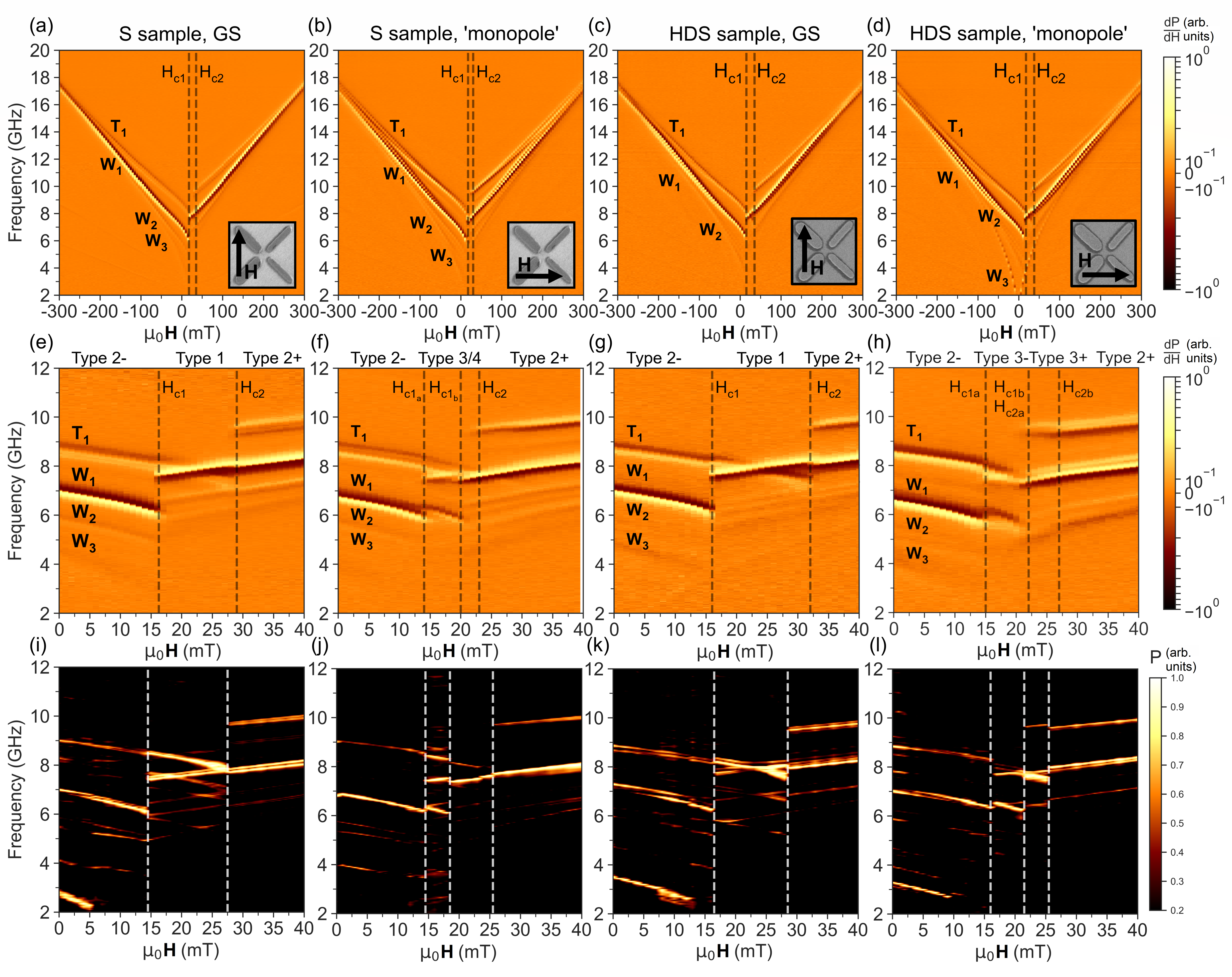}
\caption{Differential ferromagnetic resonance spectra of square (S) and high-density square (HDS) samples with corresponding micromagnetic simulations. Peak amplitude occurs at boundary between white and black bands.  \\ 
Samples were saturated in 1000 mT negative field then swept in positive field direction, with relative field orientation indicated in inset scanning electron micrographs. Measurements were performed at room temperature. \\
Fields were swept from $\pm300~$ mT, with full sweeps (a-d), 0-40 mT sweeps around the coercive fields (e-h) and micromagnetic MuMax3 simulations of the coercive field region (i-l) presented. Sample geometry and $\mathbf{H}_{ext}$ orientation are shown inset. Switching fields are labelled by vertical dashed lines, a and b subscripts refer to separate $\pm45$ and $\mp45$ subset reversal where applicable, type 3- and 3+ refer to thin-bar majority and wide-bar majority type 3 states respectively.
From left to right, vertical columns of spectra relate to samples: S (`ground-state' orientation), S (`monopole' orientation), HDS (`ground-state' orientation), HDS (`monopole' orientation).} 
\label{Fig2} \vspace{-1em}
\end{figure*}

Broadband FMR spectra were measured using a flip-chip method with samples excited by a coplanar waveguide. Frequency resolution is 20 MHz. For S and HDS samples, spectra were taken with $\mathbf{H}_{ext}$ in $\hat{y}$ (`ground-state' orientation, as in fig. \ref{Fig1} a) and $\hat{x}$ (`monopole' orientation). Samples were saturated at $\mathbf{H}_{ext} = -1000~$ mT then swept from -300 mT to 300 mT, above the saturation fields observed in MOKE data to examine spectral effects when shape anisotropy is overcome. Accompanying micromagnetic simulations were performed using MuMax3. 

Figure \ref{Fig2} shows differential FMR spectra for each sample and orientation at $\pm300~$ mT field range (fig. \ref{Fig2} a-e, relative $\mathbf{H}_{ext}$ orientation inset) and $0-40~$ mT range (fig. \ref{Fig2} f-j) with corresponding simulated spectra (fig. \ref{Fig2} k-o). 

Spectra exhibit two main Kittel-like modes, the lower and higher frequency modes corresponding to bar-centre localised modes in the wide $W_{1}$ and thin-bars $T_{1}$ respectively. This correspondence is evidenced by frequency jumps and $\frac{\partial f}{\partial \mathbf{H}}$ sign inversions indicating bar reversal\cite{gurevich1996magnetization} in the low- and high-$f$ modes at $\mathbf{H}_{c1}$ and $\mathbf{H}_{c2}$ respectively, matching switching fields observed via MOKE (fig. \ref{Fig1} c-f). Higher relative amplitude of the low-$f$ mode matches the larger sample volume share of the wide-bar, simulated spatial mode-power maps support the mode-bar correspondence. The wide-bar exhibits two higher-index modes W2 and W3, occurring at lower frequency relative to W1 due to the backward-volume wave nature of the modes. Simulated spatial mode profiles are shown in supplementary figure 1. Higher-order modes are expected in the thin-bar and observed in simulation, but fall below the amplitude threshold for experimental detection. In addition to offering two well-defined frequency channels, the different width bar subsets allow clear identification of which subset has reversed or undergone microstate-dependent frequency shifting.

For $\mathbf{H}_{ext} >$ 200 mT, thin and wide-bar modes tend to the same frequency as shape-anisotropy is overcome and bar magnetisation rotates from an Ising-like state to lie parallel to $\mathbf{H}_{ext}$. At these $\mathbf{H}_{ext}$, the bar demagnetising fields become negligble and the Kittel equation is dominated by $\mathbf{H}_{ext}$.

At lower fields around $\mathbf{H}_{c1}$ and $\mathbf{H}_{c2}$, rich and distinct spectra are observed between samples and orientations. Figure \ref{Fig2} e) shows S sample spectra in `ground-state' orientation. At $\mathbf{H}_{ext} = 0~$ the system is in a negatively-saturated type 2 microstate (fig. \ref{Fig1} h,l), the $M_x$ component of all bars oriented against positive $\mathbf{H}_{ext}$ and both modes exhibiting negative $\frac{\partial f}{\partial \mathbf{H}}$. At $\mathbf{H}_{c1} = 16~$ mT the wide-bars reverse, its mode jumping 6.2-7.8 GHz and displaying positive $\frac{\partial f}{\partial \mathbf{H}}$. The thin-bar mode is blueshifted 0.1 GHz at $\mathbf{H}_{c1}$ due to the change in local dipolar field landscape as the system enters a type 1 microstate (fig. \ref{Fig1} g,k). For $\mathbf{H}_{c1} < \mathbf{H}_{ext} < \mathbf{H}_{c2}$ the system is in a type 1 state, the two modes exhibiting opposite $\frac{\partial f}{\partial \mathbf{H}}$ and crossing at $\mathbf{H}_{ext} = 23 mT$. Opposing frequency gradients and presence of a mode crossing in this field range afford sensitive mode and gap tunability via $\mathbf{H}_{ext}$. At $\mathbf{H}_{c2} = 29~$ mT the thin-bars reverse, preparing a type 2 state aligned with $\mathbf{H}_{ext}$ and redshifting the wide-bar mode 0.1 GHz via the shift in dipolar field landscape. Above $\mathbf{H}_{c2}$ both modes exhibit positive $\frac{\partial f}{\partial \mathbf{H}}$.

Rotating $\mathbf{H}_{ext}$ $90^{\circ}$ accesses the `monopole' orientation. Fig. \ref{Fig2} b) shows that at high-field saturated states, `monopole' and `ground-state' orientations behave similarly. Around the coercive fields, fig. \ref{Fig2} f) shows stark spectral differences between the orientations.
While the `ground-state' orientation transitions directly between a type 2 and type 1 state via simultaneous reversal of both wide-bars, the highly-unfavourable type 4 dipolar field landscape separates the wide-bar switching into two distinct reversal events occurring at different fields. At $\mathbf{H}_{ext} = 13~$ mT the $\pm45^{\circ}$ subset of wide-bars better aligned to $\mathbf{H}_{ext}$ reverses, placing the system in a `thin-bar majority' type 3 state (fig. \ref{Fig1} i,m) where both thin-bars and a single wide-bar share like-polarity charges. This splits the low frequency mode as half the wide-bars ($\pm45^{\circ}$) reverse while the rest ($\mp45^{\circ}$) remain aligned against $\mathbf{H}_{ext}$. The reversed wide-bar mode jumps from 6.0-7.7 GHz and exhibits positive $\frac{\partial f}{\partial \mathbf{H}}$.
The unswitched wide-bar mode is blueshifted 0.3 GHz and the thin-bar mode redshifted 0.1 GHz by the type 3 dipolar field landscape. This reduces the gap between unswitched wide and thin-bar modes by 0.4 GHz without modifying the magnetisation state of either bar, demonstrating the degree of spectral control available via microstate engineering. Remaining wide-bars reverse at 20 mT, placing the system in a type 4 microstate (\ref{Fig1} j,n) and unifying the wide-bars in a single 7.5 GHz mode, redshifted 0.1 GHz relative to the already-reversed wide-bar mode. Thin and wide-bar modes now occupy the same frequency, obscuring the expected thin mode blueshift in the experimental data. The mode gap width may be modified by varying the relative bar widths at the fabrication stage, and the overlap here is a consequence of the specific dimensions employed. At $\mathbf{H}_{c2} =~$ 23 mT the thin-bar reverses, placing the system in a type 2 state aligned with $\mathbf{H}_{ext}$ and restoring the mode gap.

\begin{figure*}[htbp]   
\centering
\includegraphics[width=0.95\textwidth]{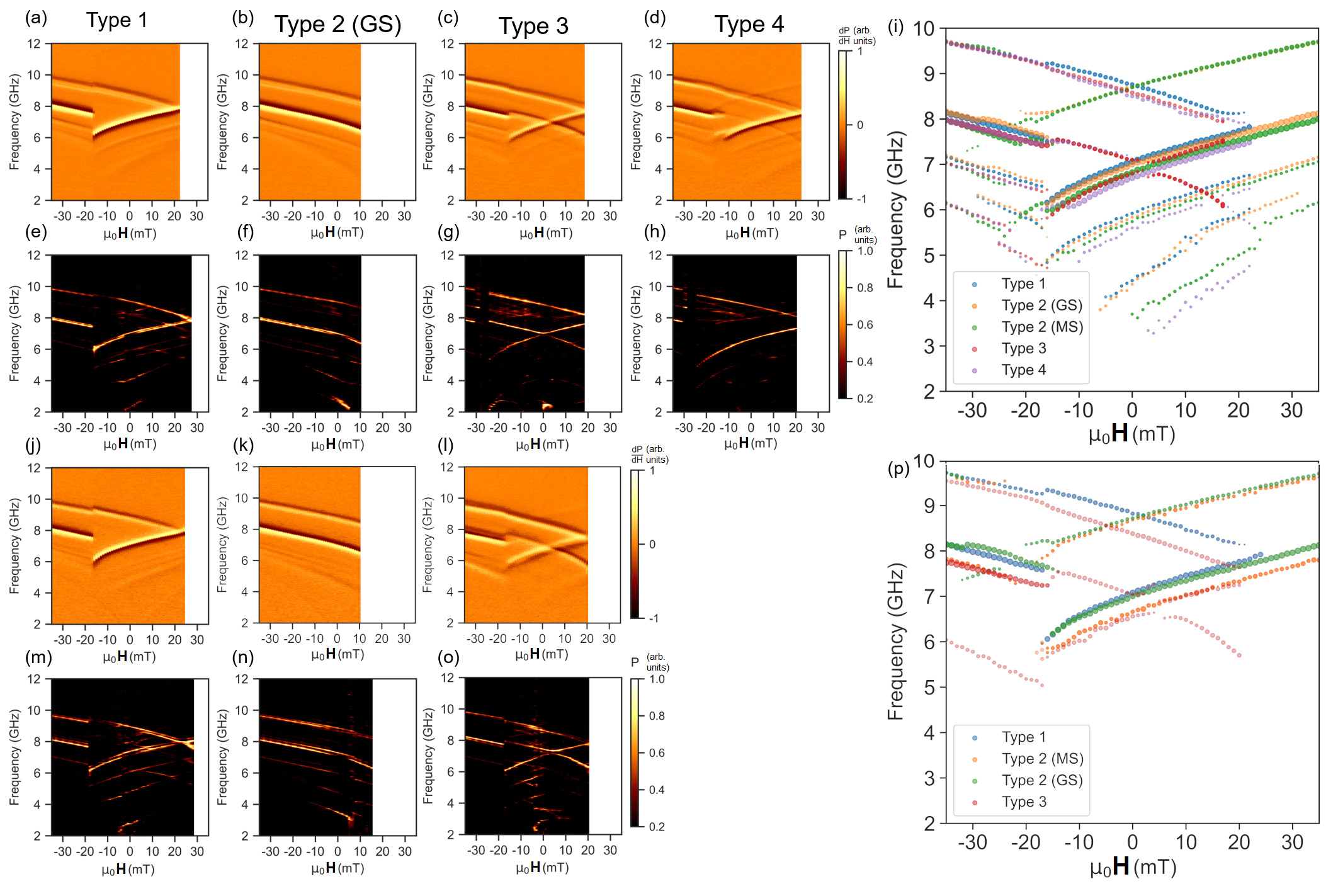}
\caption{Differential ferromagnetic resonance spectra taken while negatively sweeping $\mathbf{H}_{ext}$ after microstate preparation in positive field. Microstates were prepared by -1000 mT saturation then applying the positive field required to reverse the desired bars, hence differing positive field limits for different microstate spectra.\\
a-d) Experimental spectra for S sample microstates taken in `ground-state' (a,b) and `monopole' (c,d) orientations.\\
e-h) Simulated S sample microstate spectra for `ground-state' (e,f) and `monopole' (g,h) orientations.\\
i) Mode peak-extractions for all S sample microstate-spectra.\\
j-l) Experimental spectra for HDS sample microstates taken in `ground-state' (j,k) and `monopole' (l) orientations. `Monopole' orientation signal-to-noise is lower due to array-waveguide alignment issues. Modes are still well-resolved and correspond well with simulation.\\
m-o) Simulated HDS sample microstate spectra for `ground-state' (m,n) and `monopole' (o) orientations.\\
p) Mode peak-extractions for all HDS sample microstate-spectra.} 
\label{Fig3} \vspace{-1em}
\end{figure*}

To demonstrate the spectral control available via array design choices, the HDS sample was fabricated. By reducing bar-length and inter-bar separation, stronger dipolar interactions between bars and greater variation in the dipolar field landscape is achieved, resulting in larger spectral shifts when transitioning state. Fig. \ref{Fig2} g) shows the HDS `ground-state' orientation, qualitatively matching that of the S sample but with frequency shifts of 0.2 GHz (twice that observed in the S sample) and a broadened type 1 field window due to the increased stability. Fig. \ref{Fig2} h) shows the HDS `monopole' orientation, again qualitatively matching that of the S sample (with enhanced 0.3 GHz frequency shifts) up to 23 mT where the system transitions from a thin-bar majority type 3 state to a wide-bar majority type 3 rather than type 4. A wide-bar majority type 3 persists between 23-30 mT. At 30 mT the remaining $\pm45^{\circ}$ thin-bar reverses, causing a 0.3 GHz redshift in the thin-bar mode and transitioning to a type 2 state.

\subsection*{Negative field evolution of microstate-dependent magnonic spectra}

So far spectra have been measured while positively sweeping $\mathbf{H}_{ext}$ after negative saturation. This allows study of the system as it evolves through a range of microstates, but each microstate is stable in a limited field window. Alternatively, states may be prepared via negatively saturating then applying a microstate-specific positive field (i.e. 22 mT for the S sample type 1 state, fig. \ref{Fig2} e), then recording spectra while negatively sweeping $\mathbf{H}_{ext}$ until saturation. This allows mode dynamics to be studied for each microstate over its entire stable field range, revealing additional spectral details not accessed in fig. \ref{Fig2} e-h) 0-40 mT sweeps. Field protocols correspond to the orange traces on the figure 1 c,d,e,f) MOKE loops.

Figure \ref{Fig3} shows negatively-swept spectra for the S (FMR panels a-d), simulations e-h) and HDS samples (FMR j-m), simulations n-q) for all microstates and orientations alongside mode extractions (S sample i), HDS r) allowing state comparison. Figure \ref{Fig3} a) shows the S sample `ground-state' orientation prepared in a type 1 state at $\mathbf{H}_{ext} = 22$ mT. At 22 mT thin and wide-bar modes occupy a single frequency at 8 GHz. As field is negatively swept, modes exhibit opposite gradient due to opposing wide and thin bar magnetisation, reaching a maximum mode-frequency gap of 3 GHz at -16 mT, after which the wide-bars reverse. This prepares a type 2 state, with a wide-bar frequency jump and thin-bar redshift as observed in the 0-40 mT positive sweeps. The broadly tunable 0-3 GHz gap and wide field-stability window of type 1 state are desirable for functional magnonic systems where mode-frequency gap control is crucial.

\begin{figure}[htbp]
\centering
\includegraphics[width=0.99\textwidth]{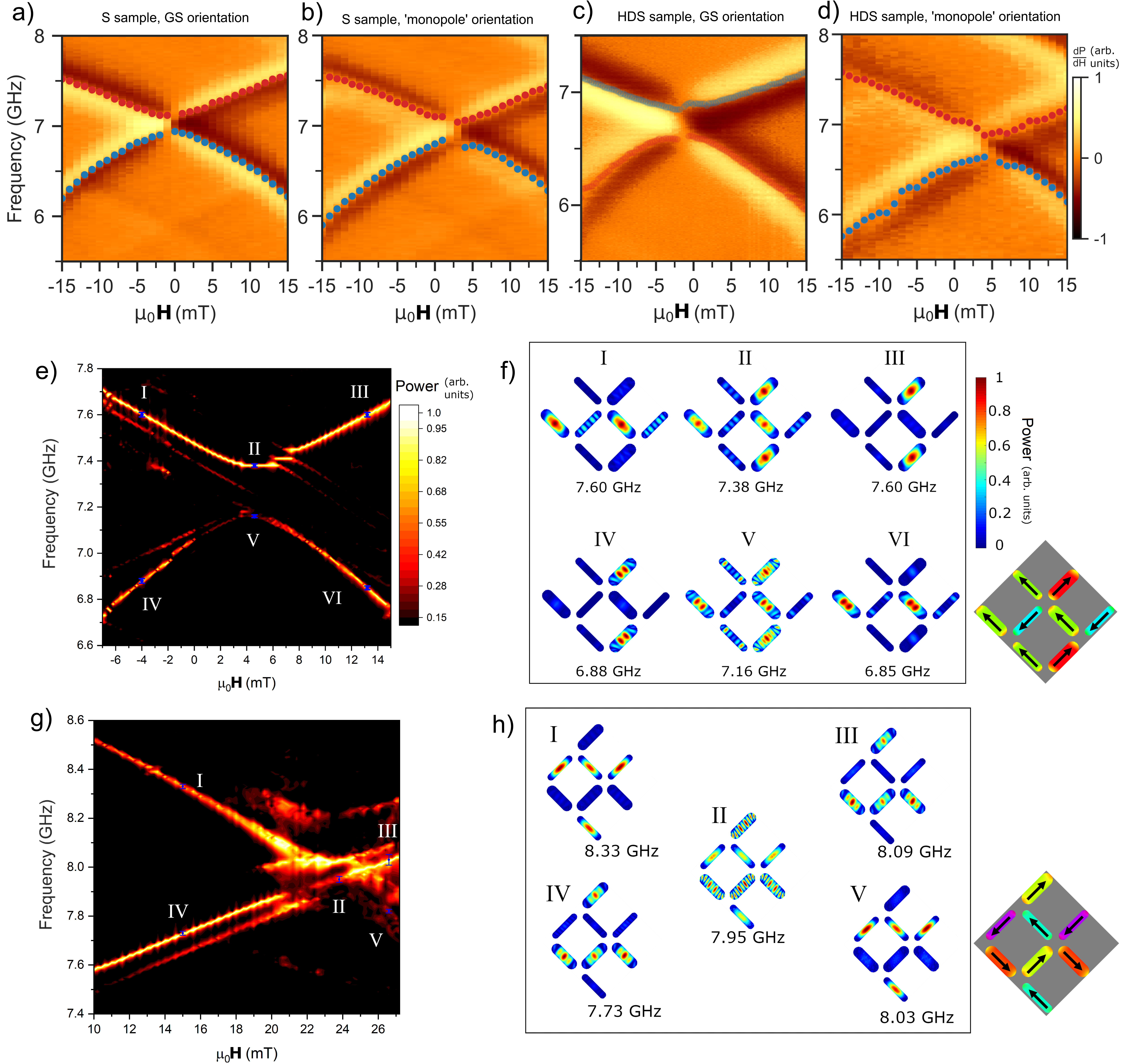}
\caption{Mode-hybridisation and anticrossings. \\
a-d) Negative-swept field dependent experimental FMR spectra of type 3 states for S and HDS samples in ground-state and `monopole' orientations. Red scatter points are peak extractions of the upper mode branch, blue points extractions of the lower branch. Mode frequency gaps or anticrossings and mode bending in the field range around the crossing point are observed in all samples and orientations. Monopole-orientation crossing points (panels b and d) are offset in positive $\mathbf{H}_{ext}$ due to the net $\mathbf{H}_{loc}$ at the vertex, ground state orientation crossings (a and c) are offset at lower magnitude, negative $\mathbf{H}_{ext}$ due to the different $\mathbf{H}_{loc}$ profile along this axis. \\
e) Simulated spectra of monopole-orientation HDS sample. Anticrossing gap of 0.32 GHz is observed at 4.3 mT. Error bars correspond to frequency range integrated over to produce spatial mode plots shown in f).\\
f) Simulated spatial mode-power maps for monopole-orientation HDS sample. Maps I-VI relate to corresponding points labelled on spectra in panel e). High-frequency, single-node mode branch is seen in I-III. Low-frequency, multi-node branch is IV-VI.
g) Simulated type 1 spectra of ground-state orientation HDS sample. Crossing occurs at 24 mT with no observable gap. Error bars correspond to frequency range integrated over to produce spatial mode plots shown in h).\\
f) Simulated spatial mode-power maps for ground-state orientation type 1 HDS sample. Maps I-V relate to corresponding points labelled on spectra in panel e). Mode-hybridisation is not observed, with matching profile pairs I and V, IV and III.
}
\label{Fig4} 
\end{figure}

Figure \ref{Fig3} b) shows the `ground-state' orientation S sample prepared in a type 2 state at 10 mT. Sweeping $\mathbf{H}_{ext}$ negatively, both modes exhibit a constant gradient and frequency gap of 2 GHz. Figure \ref{Fig3} c) shows a monopole-orientation type 3 spectra. A 21 mT preparation field reverses half the wide-bars, preparing a thin-bar majority type 3 state. The preparation here is distinct from earlier discussion of type 3 states in the HDS sample, where a well-defined $\pm45^{\circ}$ subset reverses due to closer alignment to $\mathbf{H}_{ext}$. Here, the Gaussian spread\cite{ladak2010direct} of $\mathbf{H}_{c1}$ throughout the system due to nanofabrication imperfections (termed quenched disorder\cite{libal2009creating,ladak2010direct,budrikis2012network}) is leveraged for state-preparation. By selecting a 21 mT field at the centre of the $\mathbf{H}_{c1}$ distribution, half the wide-bars are reversed and on average the system placed in a type 3 state, with a random distribution of $\pm45^{\circ}$ wide-bars reversed. While sweeping field back from 21 mT the thin-bar exhibits negative $\frac{\partial f}{\partial \mathbf{H}}$. The wide-bar mode is split into reversed and unreversed modes exhibiting opposite $\frac{\partial f}{\partial \mathbf{H}}$ sign. The two modes should cross at 5 mT if no deviation from Kittel-like behaviour is observed. However, the modes are bent away from each other around 5 mT with an anticrossing frequency gap remaining between them. The gap is observed in both experimental and simulated (fig. \ref{Fig3} g) spectra with 0.27 GHz width, and a corresponding 0.30 GHz gap in the HDS type 3 spectra (experimental and simulated in fig. \ref{Fig3} l and p respectively). Whereas previously discussed mode-frequency shifting occurs due to magnetostatic inter-bar interactions, i.e. the microstate-dependent dipolar field landscapes giving different $\mathbf{H}_{loc}$ values for the Kittel equation, mode anticrossings are an effect of dynamic mode-hybridisation\cite{kalinikos1986theory,tacchi2012forbidden,sud2020tunable,shiota2020tunable,topp2010making,iacocca2016reconfigurable}. High-resolution anticrossing spectra are shown in figure \ref{Fig4} with accompanying discussion below. In addition to the anticrossing the type 3 state offers a high-degree of spectral control, with 3 active modes and tunable mode-gaps.

Figure \ref{Fig3} d) shows the type 4 state, prepared at 22 mT. Qualitatively the spectra resembles that of the type 1 state but modes exhibit enhanced separation due to different local dipolar field landscape and are redshifted relative to type 1. This is best visualised through peak extractions shown in fig. \ref{Fig3} i). 0.4 and 0.2 GHz mode gaps at 22 mT are observed for type 4 and type 1 states respectively, along with a 0.35 GHz blueshift of the type 1 wide mode relative to type 4 demonstrating the fine control available. At -12 mT one $\pm45^{\circ}$ subset of wide-bars reverses, preparing a type 3 state with accompanying mode shifts and wide-bar mode-splitting. The remaining wide subset reverses at -16 mT, preparing a type 2 state.

The HDS sample exhibits qualitatively similar behaviour to the S sample with increased magnitude microstate-dependent frequency shifts due to stronger inter-element interaction. 
Figure \ref{Fig3} j) shows the `ground-state' orientation HDS sample prepared in a type 1 state at 30 mT. Here a 0-3.6 GHz mode gap is observed over a -12 - 30 mT field range. 
Figure \ref{Fig3} k) shows a type 2 state prepared at 16 mT, exhibiting a constant 2 GHz gap across the field sweep. 
Figure \ref{Fig3} l) shows the system in a thin-bar majority type 3 state at 21 mT. The reversed and unreversed wide-bar modes exhibit opposite gradient with a 7.3 GHz crossing at 6 mT.

\subsection*{Microstate-dependent mode hybridisation and anticrossings}

In the type 3 spectra (fig. \ref{Fig3} c),g),l) and p), where reversed and unreversed wide-bar modes approach a single frequency they do not overlap. The modes instead bend away from a Kittel-like form as they approach, leaving an anticrossing  gap\cite{macneill2019gigahertz,sud2020tunable,yuan2020enhancement,tacchi2012forbidden,iacocca2016reconfigurable} which has been predicted to occur in ASI due to a microstate-dependent band structure\cite{iacocca2016reconfigurable}. Figure \ref{Fig4} shows the anticrossing region in type 3 microstates. Samples were prepared in type 3 states in ground-state and monopole orientations, spectra measured while negatively-sweeping field. Mode-bending and anticrossings are exhibited in all experimental spectra (fig. \ref{Fig4} a-d) and corresponding simulations (fig. \ref{Fig4} f-h). Experimental spectra show anticrossing gaps of $\Delta_{S,GS}=0.27~$ GHz, $\Delta_{S,monopole}=0.27~$GHz, $\Delta_{HDS,GS}=0.22~$ GHz, $\Delta_{HDS,monopole}=0.3~$ GHz. 

Simulated spatial mode-power maps (fig. \ref{Fig4} f) show this effect occurs due to mode-hybridisation between reversed and unreversed wide-bars, causing the two modes to act like distinct upper and lower frequency v-shaped branches (red and blue peak extraction points respectively in experimental spectra) rather than diagonally intersecting Kittel-like modes. The upper-branch mode profile shows a single-node bulk mode, localised in the reversed bar at fields below the crossing point and the unreversed bar at fields above the crossing. The lower-branch profile shows a double-node bulk mode, localised in the unreversed bar at fields below crossing and vice-versa.
At the crossing point, the single-node bulk mode appears in both wide-bars at the upper-branch frequency and the multi-node mode appears in both wide-bars at the lower-branch frequency. The field at which anticrossings occur may be modified by rotating the sample between ground-state and monopole-orientations. The two orientations have different net $\mathbf{H}_{loc}$ values along the $\mathbf{H}_{ext}$-axis due to broken microstate symmetry and resultingly crossings occur at different fields, ground-state at -1 mT (-1 mT), `monopole' at 3 mT (5 mT) in S (HDS) samples. The antiferromagnetic macrospin ordering in type 3 states is crucial for mode-hybridisation. The difficulty in preparing such antiferromagnetic states is a key barrier to observing dynamic coupling effects such as anticrossings, and a key strength of the microstate-access protocol presented here.

The microstate control demonstrated allows tailoring of spectra such that modes may also cross with no resolvable anticrossing. In type 1 states (fig. \ref{Fig1} g,k), crossings are observed between thin and wide-bar modes with no observable gap or deviation from Kittel-like behaviour. Simulated spectra of the crossing point (fig. \ref{Fig4} g) show no anticrossing gap and spatial power maps (fig. \ref{Fig4} h) show a single-node bulk mode throughout the type 1 field range. While the type 1 state exhibits antiferromagnetic order between the thin and wide bars, it occurs at weaker effective interaction than type 3 states as the wide-thin bar vertex separation and dipolar-coupling are reduced relative to the type 3 wide-wide bar case. The lack of a resolvable type 1 anticrossing is testament to the sensitivity of dynamic coupling phenomena to interaction strength. Supplementary figure 2 shows simulated 0-10 GHz spectra of the HDS sample (`monopole' orientation) in a type 3 state with 0-3 GHz edge modes present and spatial magnetisation profiles of the nanoisland edges. Realignments of the static edge magnetisation occur at -1.5 mT and -6.5 mT for the wide and thin bars respectively, while the anticrossing occurs at 5 mT. As such, static magnetisation realignments are unlikely to be involved in the observed mode-hybridisation.

We have demonstrated that introducing a width-modified sublattice to ASI permits rapid, scalable and reconfigurable control over rich and diverse spectral features. This approach offers an attractive addition to the host of spectral and microstate control methodologies, requiring only widely-available global-field and nanofabrication protocols. The state-dependent spectra observed suggest microwave-assisted state preparation\cite{thirion2003switching,podbielski2007microwave,nembach2007microwave,bhat2020magnon} as a promising direction for integrated read-write functionality.

The magnitude and diversity of microstate-dependent mode and gap control exhibited invite a host of functional applications including tunable microwave filters and enable further study of how spin-wave characteristics and band structure of nanomagnetic systems may be employed in magnonic logic\cite{chumak2017magnonic} and neuromorphic devices\cite{grollier2020neuromorphic}. In particular, the observation of previously elusive microstate-dependent mode-hybridisation and anticrossings suggests a magnonic device which may reconfigurably transmit or reflect spin-waves depending on its state. In this regard we emphasise that anticrossing behaviour depends only on the microstate and does not require width-modification except as a means for microstate access. As state-preparation techniques develop, we expect mode-hybridisation to become observable and exploitable in other artificial spin systems.

\section*{Methods}

Simulations were performed using MuMax3. To maintain field sweep history, ground state files are generated in a separate script and used as inputs for dynamic simulations. S sample dimensions are; wide: 800 by 230 by 20 nm, narrow: 800 by 130 by 20 nm and lattice parameter: 1120 nm (gap = 160 nm). HDS sample dimensions are; wide: 600 by 200 by 20 nm and narrow: 600 by 130 by 20 nm and lattice parameter: 800 nm (gap = 100 mn). Perturbation of dimensions from SEM images were introduced to more accurately reproduce both static and dynamic magnetisation behaviour. Material parameters for NiFe used are; saturation magnetisation, M$_{sat}$ = 750 kA/m, exchange stiffness. A$_{ex}$ = 13 pJ and damping, $\alpha$ = 0.001  All simulations are discretized with lateral dimensions, c$_{x,y}$ = 5 nm and normal direction, c$_z$ = 10 nm and periodic boundary conditions applied to generate lattice from unit cell. A broadband field excitation sinc pulse function is applied along z-direction with cutoff frequency = 20 GHz, amplitude = 0.5 mT. Simulation is run for 25 ns saving magnetisation every 25 ps. Static relaxed magnetisation at t = 0 is subtracted from all subsequent files to retain only dynamic components, which are then subject to a FFT along the time axis to generate a frequency spectra. Power spectra across the field range are collated and plotted as a colour contour plot with resolution; $\Delta f$ = 40 MHz and $\Delta \mu_0 H$ = 1 mT. Spatial power maps are generated by integrating over a range determined by the full width half maximum of peak fits and plotting each cell as a pixel whose colour corresponds to its power. Each colour plot is normalised to the cell with highest power. High-resolution simulations performed for figure \ref{Fig4} have lower damping, $\alpha = 0.0001$, and are run for 100 ns saving every 50 ps. This produces colour plots with resolution; $\Delta f$ = 10 MHz and $\Delta \mu_0 H$ = 0.2 mT. $\mathbf{H}_{ext}$ is offset from the array $\hat{x},\hat{y}$-axes by $1^{\circ}$ to better match experiment. Lowering alpha reduces mode linewidth and allows for better resolution of mode behaviour particularly when multiple modes are present in close frequency proximity, as in the anticrossing case. 

Samples were fabricated via electron-beam lithography liftoff method on a Raith eLine system with PMMA resist. Ni$_{81}$Fe$_{19}$ (permalloy) was thermally evaporated and capped with Al$_2$O$_3$. A `staircase' subset of bars was increased in width to reduce its coercive field relative to the thin subset, allowing independent subset reversal via global field.

Ferromagnetic resonance spectra were measured using a NanOsc Instruments cryoFMR in a Quantum Design Physical Properties Measurement System. Broadband FMR measurements were carried out on large area samples $(\sim 2 \times 2~ \text{ mm}^2)$ mounted flip-chip style on a coplanar waveguide. The waveguide was connected to a microwave generator, coupling RF magnetic fields to the sample. The output from waveguide was rectified using an RF-diode detector. Measurements were done in fixed in-plane field while the RF frequency was swept in 20 MHz steps. The DC field was then modulated at 490 Hz with a 0.48 mT RMS field and the diode voltage response measured via lock-in. The experimental spectra show the derivative output of the microwave signal as a function of field and frequency. The normalised differential spectra are displayed as false-colour images with symmetric log colour scale.

Magnetic force micrographs were produced on a Dimension 3100 using commercially available normal-moment MFM tips.

MOKE measurements were performed on a Durham Magneto-Optics NanoMOKE system. The laser spot is approximately 20 $\mu$m diameter. The longitudinal Kerr signal was normalised and the linear background subtracted from the saturated magnetisation. The applied field is a quasistatic sinusoidal field cycling at 11 Hz and the measured Kerr signal is averaged over 300 field loops to improve signal to noise. 



\subsection*{Author contributions}
JCG, AV, TD and WRB conceived the work.\\
JCG, KDS and AV fabricated the samples. AV performed CAD design of the structures.\\
AV performed experimental MOKE measurements and the majority of FMR measurements.\\
JCG performed FMR measurements on the HDS sample, MS orientation.\\
AV performed analysis of FMR measurements and generation of spectral heatmaps.\\
AV, JCG and KDS performed MFM measurements.\\
TD wrote code for simulation of the magnon spectra and performed micromagnetic simulations. Noticed avoided crossings in simulation prompting further investigation into experimental data.\\
DMA wrote code for simulation of the magnon spectra. \\
JCG drafted the manuscript, with contributions from all authors in editing and revision stages.\\

\subsection*{Acknowledgements}
This work was supported by the Leverhulme Trust (RPG-2017-257) to WRB. \\
TD and AV were supported by the EPSRC Centre for Doctoral Training in Advanced Characterisation of Materials (Grant No. EP/L015277/1).\\
Simulations were performed on the Imperial College London Research Computing Service\cite{hpc}.\\
The authors would like to thank Professor Lesley F. Cohen of Imperial College London for enlightening discussion and comments, and David Mack for excellent laboratory management.

\subsection*{Competing interests}
The authors declare no competing interests.

\subsection*{Data availability statement}
The datasets generated during and/or analysed during the current study are available from the corresponding author on reasonable request.

\subsection*{Code availability statement}
The code used in this study is available from the corresponding author on reasonable request.

\section*{Supplementary Information}

\begin{figure*}[thbp]   
\centering
\includegraphics[width=0.99\textwidth]{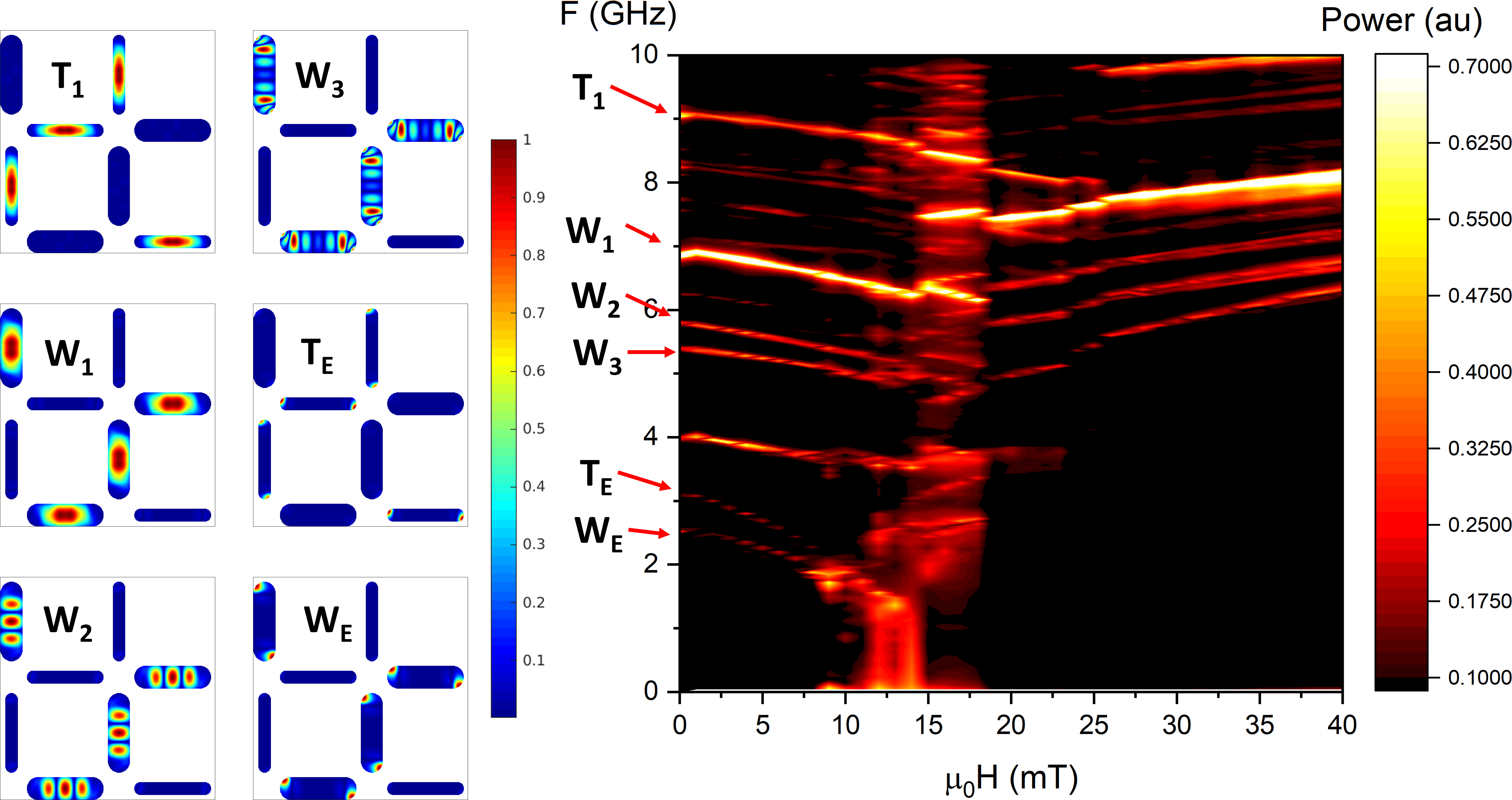}
\caption*{\textbf{Supplementary figure 1} Simulated spatial mode profiles of the S sample (`monopole' orientation) taken at zero field along with corresponding 0-40 mT spectra.} 
\label{SuppModeMaps} \vspace{-1em}
\end{figure*}

\subsection*{Supplementary note 1 - Simulated spatial mode profiles}

\begin{figure*}[thbp]   
\centering
\includegraphics[width=0.99\textwidth]{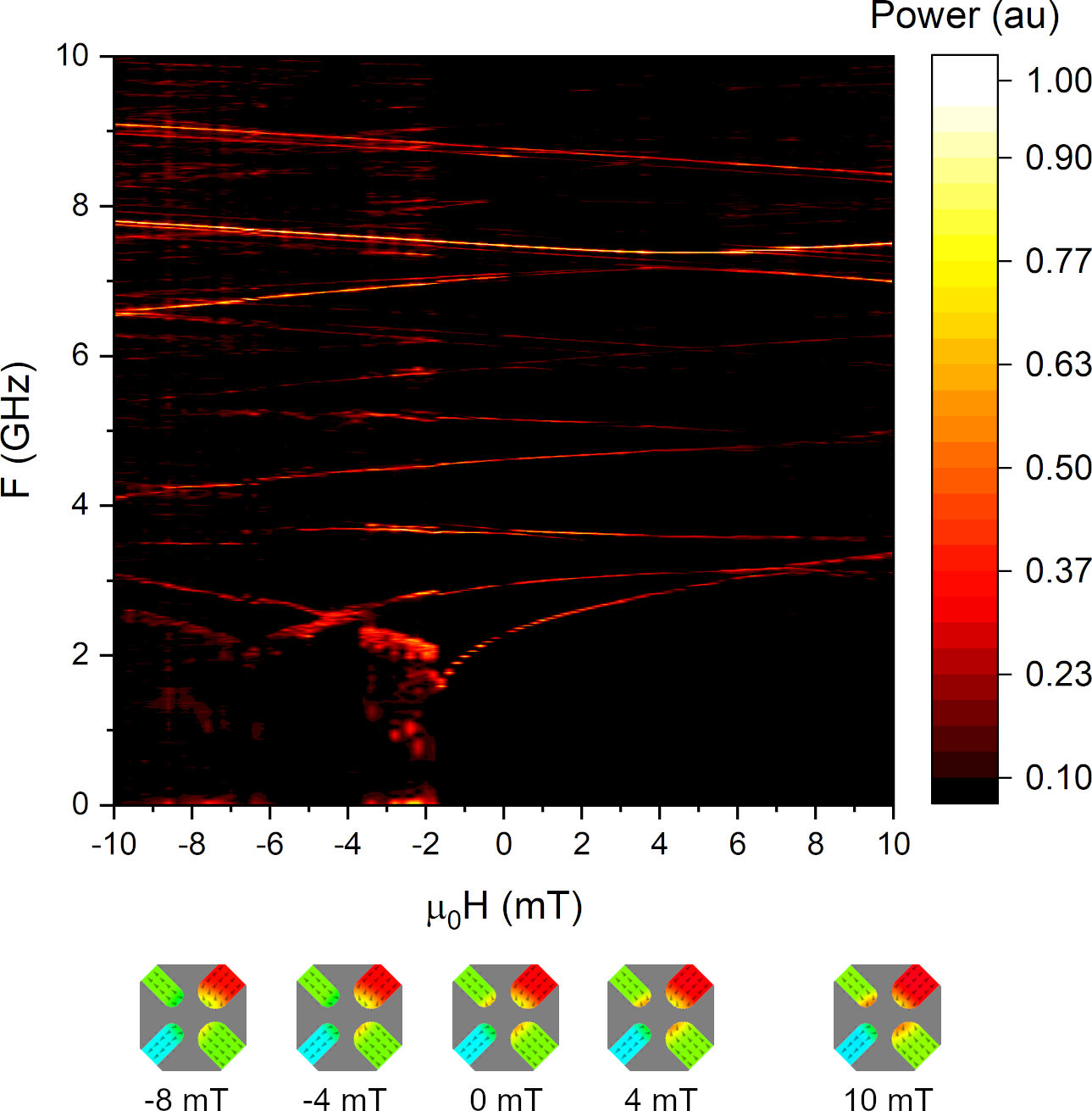}
\caption*{\textbf{Supplementary figure 2} Simulated spectra of HDS sample (`monopole' orientation) prepared in type 3 microstate. 0-10 GHz frequency range shows edge mode behaviour (0-3 GHz range) of wide and thin bars and anticrossing of wide bar bulk modes at ~5 mT, 7.2 GHz.} 
\label{expandedT3} \vspace{-1em}
\end{figure*}

Micromagnetic simulations of spatial mode profiles taken at zero field and corresponding 0-40 mT spectra for the S sample (`monopole' orientation) are shown in supplementary figure 1. Mode labelling numbers W1, W2 and W3 are sequential and do not denote mode index number.

\subsection*{Supplementary note 2 - Extended frequency and field range spectra of type 3 spectra}

Supplementary figure 2 shows simulated spectra for the HDS sample (`monopole' orientation) prepared in the type 3 state. Corresponding to an wider frequency range version of the spectra shown in figure 3 e), the edge modes of the thin and wide bars are observed in the 0-3 GHz frequency range. The wide bar edge mode reverses gradient at -1.5 mT, the thin bar edge mode reverses gradient at -6.5 mT. These gradient reversals are caused by changes in the static magnetisation curl states of the nanoisland edge regions, which are shown at -8, -4, 0, 4 and 10 mT below the spectra. These state changes occur far from the anticrossing point at 5 mT and as such, rearrangements of the static edge magnetisation states are unlikely to be linked to the mode hybridisation behaviour.

\subsection*{Supplementary note 3 - Positions of the MOKE magnetization plateaux}
Bar reversal occurs by a domain wall nucleation process at a field determined by the aspect ratios of the bars. The wide bar subset can be characterized by a mean coercive field $H_{c1}$, and a magnetization $M_{wide}$ and the thin bar subset by a mean coercive field $H_{c2}$ and a magnetization $M_{thin}$. Because the other bar dimensions are identical the ratio of the volumes and magnetisations is the same as the ratio of the bar widths. For a sample with all bars identical, the type 1 and type 4 state should have zero net magnetization. For the width modified sample the different volumes of reversed and unreversed bars would be expected (for perfect Ising spins and nominal bar widths) to give $M/M_{type 2}$ = $\frac{M_{wide}-M_{thin}}{M_{wide}+M_{thin}} = \frac{t_{wide}-t_{thin}}{t_{wide}+t_{thin}}~$ = 0.226 for the S sample and 0.231 for the HDS sample (for both type 1 and type 4). The relative magnetization of the ground-state minor loop to the saturated major loop at zero field is approx. 0.3 in the S sample and 0.2 in the HDS sample. For the type 4 state it is approx. 0.5 and 0.6 for S and HDS respectively. In a type 3 that is formed by both wide bars switching and triggering one thin bar to also reverse, then $M/M_S$ would be $\frac{t_{wide}}{t_{wide}+t_{thin}}$ = 0.613 for S and 0.615 for HDS.
Imperfections in the nanofabrication (quenched disorder) give the sublattice switching fields a Gaussian distribution about the mean, with a standard deviations $\sigma_{wide}$ around $H_{c1}$ and $\sigma_{thin}$ around $H_{c2}$. If bars were all sufficiently spaced to be not interacting then desired states could be accessed by applying $H_{ext} = H_{c1}+H_{c2}/2$ as long as $H_{c2}-H_{c1} >> \sigma_{wide}  + \sigma_{thin}$.
However we are in the strongly-interacting regime and so each bar experiences an effective field $H_{eff} = H_{app}+H_{loc}$. The reversal of wide bars will change $H_{loc}$ experienced by the thin bars. If $\Delta H_{loc}$ increases $H_{eff}$ then this makes it more difficult to realise the ordered state, as do the cases where we are preparing a type 4 state. Where we are writing the type 1 state from the saturated type 2 state, $\Delta H_{loc}$ decreases $H_{eff}$ and so the interactions increase the operating window where the ordered state may be prepared. The difference can be seen in the MOKE hysteresis loops in fig. 1 c,d,e,f). In fig. 1 c,e) we have very clear plateaux in the major (blue) hysteresis loops and can very easily and reproducibly send minor loops to the type 1 microstate and back to saturated. Note that the data is the average of thousands of individual loops and so the sharp switching and flat plateuax show there is no significant stochasticity in this major hysteresis loop and we go through the same microstates at the same fields in each measurement. Similarly in the minor (orange) loops we repeatedly go the the same expected plateau magnetization. For the same sample, with the same extrinsic disorder and sigmas, in the monopole geometry, switching the wide bars causes the dipolar field of all neighbouring bars to help the reversal of the thin bars and so the two Gaussian distributions start to overlap. We know from MFM we can access large areas of pure type 4 with the correct protocol, but the hysteresis loop shows very broad reversal with no clear plateaux. It is not clear from our data whether the broadening we see is from averaging similar broad loops or different loops with sharper individual features. The disorder could be spatial within the measurement spot, temporal with loop cycle number or both. Certainly there is a significant stochastic contribution in the measurement.

\end{document}